\documentclass[aps,prl,reprint,amsmath,amssymb,groupedaddress,nobibnotes,showpacs]{revtex4-1}

\usepackage{graphicx}
\usepackage{dsfont}

\begin{document}

\title{Experimental verification of the commutation relation for Pauli spin operators using single-photon quantum interference}

\author{Yong-Su Kim}\email{yskim25@gmail.com}
\affiliation{Department of Physics, Pohang University of Science and
Technology (POSTECH), Pohang, 790-784, Korea}

\author{Hyang-Tag Lim}
\affiliation{Department of Physics, Pohang University of Science and
Technology (POSTECH), Pohang, 790-784, Korea}

\author{Young-Sik Ra}
\affiliation{Department of Physics, Pohang University of Science and
Technology (POSTECH), Pohang, 790-784, Korea}

\author{Yoon-Ho Kim} \email{yoonho72@gmail.com}
\affiliation{Department of Physics, Pohang University of Science and
Technology (POSTECH), Pohang, 790-784, Korea}

\date{\today}

\begin{abstract}
We report experimental verification of the commutation relation for Pauli spin operators using quantum interference of the single-photon polarization state. By superposing the quantum operations $\sigma_z \sigma_x$ and $\sigma_x \sigma_z$ on a single-photon polarization state, we have experimentally implemented the commutator,  $[\sigma_{z}, \sigma_{x}]$, and the anticommutator, $\{\sigma_{z}, \sigma_{x}\}$, and have demonstrated the relative phase factor of $\pi$ between $\sigma_z \sigma_x$ and $\sigma_x \sigma_z$ operations. The experimental quantum operation corresponding to the commutator, $[\sigma_{z}, \sigma_{x}]=k\sigma_y$, showed process fidelity of 0.94 compared to the ideal $\sigma_y$ operation and $|k|$ is determined to be $2.12\pm0.18$.
 \end{abstract}


\maketitle


One of the most fundamental differences between classical and quantum physics is that, in quantum physics, a physical quantity always corresponds to a specific Hermitian operator or an observable and certain pairs of observables are non-commuting. In quantum physics, as a result, physical quantities that correspond to non-commuting observables `\textit{cannot have simultaneous reality}' \cite{epr}. As precise knowledge of one physical quantity precludes precise knowledge of the other physical quantity for a pair of non-commuting observables, the commutation relation is also at the very heart of the Heisenberg uncertainty relation \cite{heisenberg27, robertson29}. For example, it is known that the position, $\hat X$, and the momentum, $\hat P$, operators of a quantum mechanical particle do not commute, $[\hat{X}, \hat{P}]=\hat{X}\hat{P}-\hat{P}\hat{X}=i\hbar$, and this leads to the position-momentum uncertainty relation, $\triangle{x}\triangle{p} \geq {\hbar}/{2}$, which can be beautifully demonstrated in a single-slit diffraction experiment \cite{feynman}. It is also interesting to note that the position-momentum uncertainty principle, which is derived from the commutation relation, has long been believed to enforce Bohr's complementarity in a two-slit experiment \cite{bohr}, although recent research on delayed choice quantum erasure ruled out this possibility \cite{kim00}. 


Although the commutation relation has been very well understood theoretically, direct experimental verification of the commutation relation for a pair of conjugate observables has been rather limited. For fermions, Pauli anti-commutation has been demonstrated in neutron polarimetry and interferometry \cite{wagh97,hasegawa97}. For bosons, the commutator, $[\hat{a}, \hat{a}^{\dag}]$,  and the anticommutator, $\{\hat{a}, \hat{a}^{\dag}\}$, have been demonstrated for the photon creation, $\hat{a}^{\dag}$, and annihilation, $\hat{a}$, operators, implemented with post-selective single-photon addition and subtraction operations, respectively  \cite{kim08, zavatta09}.

In this paper, we report an experimental verification of the commutation relation for Pauli spin operators using quantum interference of the single-photon polarization state. By combining the bosonic interference property of a single-photon and Pauli spin operators for the single-photon polarization state, we have observed the relative phase factor of $\pi$ between $\sigma_z \sigma_x$ and $\sigma_x \sigma_z$ operations and have experimentally verified the quantum operations corresponding to the commutator, $[\sigma_{z}, \sigma_{x}]$, and the anticommutator, $\{\sigma_{z}, \sigma_{x}\}$, by means of quantum process tomography.


We begin by introducing the basic idea behind the experiment. The Pauli spin operators are defined as,
\begin{equation}\label{Pauli_matrices}
\sigma_{x}=\left(\begin{matrix}
           0 & 1 \\
           1 & 0 \\
         \end{matrix}\right),
\  \sigma_{y}=\left(\begin{matrix}
           0 & -i \\
           i & 0 \\
         \end{matrix}\right),
\  \sigma_{z}=\left(\begin{matrix}
           1 & 0 \\
           0 & -1 \\
         \end{matrix}\right),
\end{equation}
and they are essential in describing the unitary evolution (i.e., rotation) and the projection measurement of a two-level quantum system (i.e., a qubit), such as, the spin state of a spin-$1/2$ fermions (which requires the factor of $\hbar/2$), the polarization state of a single-photon, etc. 

We can observe non-commutativity of the Pauli spin operators by applying them on an arbitrary qubit  $|\psi_0\rangle = \alpha |0\rangle + \beta|1\rangle = (\begin{smallmatrix}
           \alpha \\
           \beta \\
         \end{smallmatrix})$. For example, the Pauli spin operators $\sigma_z$ and $\sigma_x$ acting on the qubit $|\psi_0\rangle$ cause the state to evolve as
\begin{eqnarray}\
\sigma_{z}\sigma_{x}|\psi_{0}\rangle=i\sigma_{y}|\psi_{0}\rangle=\left(\begin{matrix}
           \beta \\
           -\alpha \\
         \end{matrix}\right),\\
\sigma_{x}\sigma_{z}|\psi_{0}\rangle=-i\sigma_{y}|\psi_{0}\rangle=\left(\begin{matrix}
           -\beta \\
           \alpha \\
         \end{matrix}\right).
\end{eqnarray}
It can be seen in the above equations that the order in which the operators $\sigma_z$ and $\sigma_x$ are applied to the qubit $|\psi_0\rangle$ affects the outcome, showing non-commutativity of the operators. In particular, there exists the relative phase difference of $\pi$ between eq.~(2) and eq.~(3) which we shall verify experimentally in this paper.

The commutator and the anticommutator for the $\sigma_z$ and $\sigma_x$ operators are given, respectively, as
\begin{eqnarray}
[\sigma_{z}, \sigma_{x}]|\psi_0\rangle &=& (\sigma_{z}\sigma_{x}-\sigma_{x}\sigma_{z})|\psi_0\rangle=2i\sigma_{y}|\psi_0\rangle,\\
\{\sigma_{z},\sigma_{x}\}|\psi_0\rangle &=& (\sigma_{z}\sigma_{x}+\sigma_{x}\sigma_{z})|\psi_0\rangle=0.
\end{eqnarray}
Also, it is easily shown that $[\sigma_{z}, \sigma_{z}]=0$ and $\{\sigma_{z},\sigma_{z}\}=2I$. Using the above results, it is possible to show that the commutator and the anti-commutator, in general, take the form
\begin{equation}\label{comm}
[\sigma_{j}, \sigma_{k}] = 2i\epsilon_{jkl}\sigma_{l}, \,\,\, \{\sigma_{j}, \sigma_{k}\} = 2 I\delta_{jk},
\end{equation}
where the subscripts represent the cartesian coordinates $x$, $y$, and $z$, $\epsilon_{jkl}$ is the
Levi-Civita symbol, and $\delta_{jk}$ is the Kronecker delta. In this paper, we experimentally implement the quantum operations corresponding to the commutator, $[\sigma_{z}, \sigma_{x}]$, and the anti-commutator, $\{\sigma_{z}, \sigma_{x}\}$, by superposing $\sigma_z \sigma_x$ and $\sigma_x \sigma_z$ operations.

Our choice of the physical system for experimentally verifying the commutation relation for Pauli spin operators is the polarization state of a single-photon or the single-photon polarization qubit. This is a particularly convenient choice as all single-qubit quantum operations on the polarization qubit can be implemented with half-wave and quarter-wave plates and superpositions of quantum states (or operations) can be implemented with an interferometer. In particular, if we define $|0\rangle$ as the horizontal and $|1\rangle$ as the vertical polarizations, $\sigma_x$ and $\sigma_z$ operations on a polarization qubit can be realized by using a half-wave plate with its fast axis oriented at $45^\circ$ and at $0^\circ$, respectively, with respect to the vertical polarization.

\begin{figure}[t]
\includegraphics[width=3.4in]{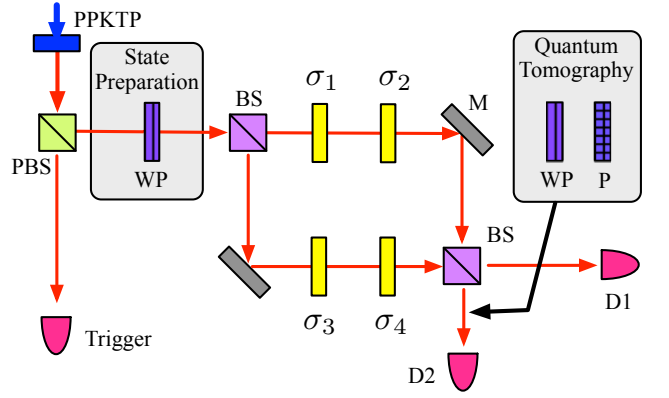}
\caption{Schematic of the experiment. The heralded single-photon source consists of a diode-laser pumped type-II PPKTP crystal, a polarizing beam splitter PBS, and a trigger detector. An arbitrary polarization state $|\psi_0\rangle$ is prepared with a set of half-wave and quarter-wave plate (WP). Pauli spin operators are realized with half-wave plates ($\sigma_1$, $\sigma_2$, $\sigma_3$, and $\sigma_4$) and the operator superposition is implemented with the single-photon Mach-Zehnder interferometer. Mirror M is used to set the relative phase difference between the two arms of the interferometer at the integer multiple of $2\pi$ so that detector D1 is located at the anti-commutator output port and detector D2 is located at the commutator output port. A set of wave plate WP and a polarizer P can be inserted to D2 output port for quantum state tomography.
}\label{setup}
\end{figure}

The experimental schematic is shown in Fig.~\ref{setup}. A type-II PPKTP crystal pumped by a multi-mode diode laser operating at 405 nm generates a pair of orthogonally polarized photons via the spontaneous parametric down-conversion process. The signal-idler photon pair is then split by the polarizing beam splitter PBS. Detection of the horizontally polarized idler photon at the trigger detector signals that the vertically polarized signal photon is prepared in the localized single-photon state \cite{hong,baek08, kim09}. A set of pump blocking dichroic mirrors and a 10 nm bandpass filter (not shown in the figure) are used to suppress the pump noise. The signal photon is then prepared in the arbitrary polarization state $|\psi_0\rangle$ with a set of zero-order half-wave and quarter-wave plates (WP). 

If the signal photon gets transmitted at the beam splitter BS, see Fig.~\ref{setup}, the polarization qubit $|\psi_0\rangle$ undergoes the state transformation, $\sigma_2 \sigma_1 |\psi_0\rangle/\sqrt{2}$. If the signal photon gets reflected at BS, the polarization qubit undergoes a different state transformation,  $i\sigma_4 \sigma_3 |\psi_0\rangle/\sqrt{2}$. Note that the Pauli spin operators $\sigma_1$,  $\sigma_2$, $\sigma_3$, and $\sigma_4$ are all implemented with zero-order half-wave plates. The two amplitudes $\sigma_2 \sigma_1 |\psi_0\rangle/\sqrt{2}$ and $i\sigma_4 \sigma_3 |\psi_0\rangle/\sqrt{2}$ can be coherently added (or superposed) if the reflected and transmitted modes are combined at the second beam splitter with the path length difference much smaller than the coherence length of the input signal photon. Thus, at the D1 output port of the interferometer, the input qubit $|\psi_0\rangle$ will be found to have undergone the quantum operation
\begin{equation}
\frac{i}{2}(\sigma_{2}\sigma_{1}e^{i\phi} + \sigma_{4}\sigma_{3}),\label{d1}
\end{equation}
and, at the D2 output, the input qubit $|\psi_0\rangle$ will be found to have undergone a different quantum operation
\begin{equation}
\frac{1}{2}(\sigma_{2}\sigma_{1}e^{i\phi} - \sigma_{4}\sigma_{3}),\label{d2}
\end{equation}
where $\phi$ is the relative phase set by the mirror M.  Finally, the coincidence events between the detectors D1/D2 and the trigger detector are measured. In front of the detector D2, a set of half-wave and quarter wave plates (WP) and a polarizer P can be inserted for quantum state tomography measurements \cite{james01}.

If the single-qubit operations are set such that $\sigma_4=\sigma_1$ and $\sigma_3=\sigma_2$ and if $\phi=2\pi n$ where $n$ is an integer, the single-photon polarization qubit found at D1 and D2 output ports will have undergone the quantum operations corresponding to the anti-commutator, $\frac{i}{2}\{\sigma_2,\sigma_1\}$, and the commutator, $\frac{1}{2}[\sigma_2,\sigma_1]$, respectively. As we are interested in implementing these operations, for the rest of the paper, the quantum operations are restricted to $\sigma_{4}=\sigma_{1}$ and
$\sigma_{3}=\sigma_{2}$.

\begin{figure}[t]
   \includegraphics[width=3.4in]{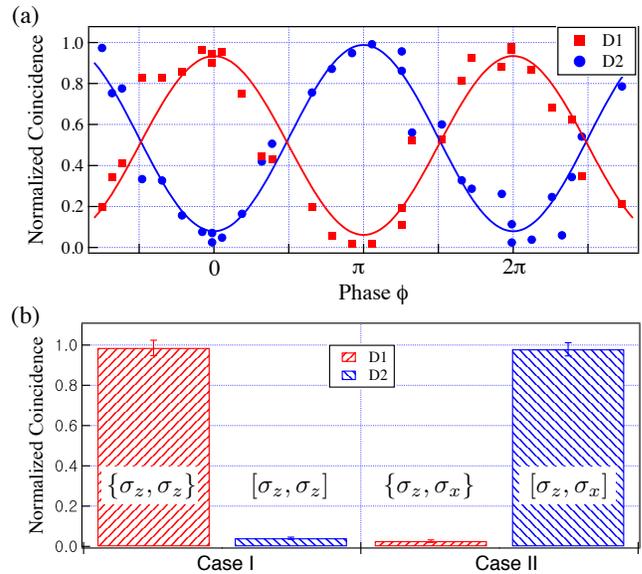}
  \caption{(a) All single-qubit operations are set at $\sigma_1=\sigma_2=\sigma_3=\sigma_4=\sigma_z$. Solid lines are sinusoidal fits to the data. 
  (b) Case I setting: when the phase is set at $\phi=0$, events at D1 corresponds to $\frac{1}{4}\langle \psi_0 | \{\sigma_z,\sigma_z\}^\dag \{\sigma_z,\sigma_z\}|\psi_0\rangle$ measurement and events at D2 corresponds to $\frac{1}{4}\langle \psi_0 | [\sigma_z,\sigma_z]^\dag [\sigma_z,\sigma_z]|\psi_0\rangle$ measurement. Case II setting:  With the phase fixed at $\phi=0$, we set $\sigma_1=\sigma_4=\sigma_x$ while keeping  $\sigma_2=\sigma_3=\sigma_z$. Events at D1 now corresponds to $\frac{1}{4}\langle \psi_0 | \{\sigma_z,\sigma_x\}^\dag \{\sigma_z,\sigma_x\}|\psi_0\rangle$ measurement and events at D2 corresponds to $\frac{1}{4}\langle \psi_0 | [\sigma_z,\sigma_x]^\dag [\sigma_z,\sigma_x]|\psi_0\rangle$ measurement. The changes in the normalized coincidence rates (from case I to case II) show that there exists the phase difference of $\pi$ between $\sigma_{z}\sigma_{x}|\psi_0\rangle$ and $\sigma_{x}\sigma_{z}|\psi_0\rangle$, i.e. $\sigma_{z}\sigma_{x}=-\sigma_{x}\sigma_{z}$.}\label{data}
\end{figure}

In the experiment, the single-qubit operations are set at either $\sigma_z$ or at $\sigma_x$ and this can be easily accomplished by rotating the angles of zero-order half-wave plates. To set the phase $\phi$ properly, we make use of the following property of the commutator, $[\Lambda,\Lambda]=0$ for any operator $\Lambda$, and when this operation is applied to the quantum state $|\psi_0\rangle$, null outcome is expected. Thus, we first set all the single-qubit quantum operations at $\sigma_1=\sigma_2=\sigma_3=\sigma_4=\sigma_z$ and this corresponds to setting the fast axes of the zero-order half-wave plates at $0^\circ$ with respect to the vertical polarization. At this condition, if the phase $\phi$ is set correctly at integer multiples of $2\pi$, the detector D2  should register no photons at all (corresponding to the measurement outcome of the $\frac{1}{2}[\sigma_z,\sigma_z]$ operation,  $\frac{1}{4}\langle \psi_0 | [\sigma_z,\sigma_z]^\dag [\sigma_z,\sigma_z]|\psi_0\rangle$) while the detector D1  registers the maximum photon count rate (corresponding to the measurement outcome of the $\frac{1}{2}\{\sigma_z,\sigma_z\}$ operation,  $\frac{1}{4}\langle \psi_0 | \{\sigma_z,\sigma_z\}^\dag \{\sigma_z,\sigma_z\}|\psi_0\rangle$). Experimentally, we scan the mirror M while observing the coincidence counts between the trigger detector and D1/D2. The mirror position is then fixed at $\phi=0$ where the detection rates of D1 is at its maximum and D2 is at its minimum. See, for example, Fig.~\ref{data}(a). (We shall  refer to this setting as \textit{case I}.) 

With the phase fixed at $\phi=0$, we now set $\sigma_1=\sigma_4=\sigma_x$ while keeping  $\sigma_2=\sigma_3=\sigma_z$. (We refer to this setting as \textit{case II}.) In experiment, this corresponds to rotating the fast axes of the zero-order half-wave plates  $\sigma_1$ and $\sigma_4$ from $0^\circ$ to $45^\circ$ with respect to the vertical polarization. It is important to make sure that the phase $\phi$ set in the previous measurement is not affected by the rotation of the wave plate. In the experiment, we isolated the interferometer setup in an enclosure and motorized the mirror and the wave plate controls to make sure that the wave plate rotation only affects the polarization state of the photon. In this setting, the detection events at D1 now corresponds to  $\frac{1}{4}\langle \psi_0 | \{\sigma_z,\sigma_x\}^\dag \{\sigma_z,\sigma_x\}|\psi_0\rangle$ measurement and the detection events at D2 corresponds to $\frac{1}{4}\langle \psi_0 | [\sigma_z,\sigma_x]^\dag [\sigma_z,\sigma_x]|\psi_0\rangle$ measurement. Due to the phase difference of $\pi$ shown in eq.~(2) and eq.~(3) between $\sigma_z \sigma_x$ and $\sigma_x \sigma_z$ operations, we now expect that D1 will register no photons but D2 will register the maximum photon count rate. 

Therefore, experimental observation of the shift of the photo-count distributions from case I (D1 at maximum rate; D2 at zero) to case II (D1 at zero; D2 at maximum rate) verifies the relation $\sigma_{z}\sigma_{x}=-\sigma_{x}\sigma_{z}$ (i.e.,  $\{\sigma_z,\sigma_x\}=0$) and allows us to experimentally construct quantum operations corresponding to the commutator and the anti-commutator for $\sigma_z$ and $\sigma_x$. Note that the commutator and the anti-commutator operations correspond to different output ports of the interferometer. The experimental data are shown in Fig.~\ref{data}(b). The changes in the normalized coincidence rates from case I to case II indeed show the expected behavior: D1 at maximum rate and D2 at zero for case I and D1 at zero and D2 at maximum rate for case II. The data, therefore, experimentally confirms the phase shift of $\pi$ in eq.~(2) and eq.~(3). It is interesting to point out that the $\pi$ phase shift of this kind cannot be observed by performing quantum state tomography.

Although the data shown in Fig.~\ref{data}(b) confirms the $\pi$ phase shift between the operations $\sigma_z \sigma_x$ and $\sigma_x \sigma_z$, we have yet to confirm the commutator relation $[\sigma_z, \sigma_x]= k \sigma_y$, where $k=2i$ theoretically. This requires quantum process tomography (QPT) which allows us to experimentally characterize the set of quantum processes operated on the input qubit  $|\psi_0\rangle$ \cite{kim09, fiurasek01}. We therefore inserted a set of quarter-wave and half-wave plates (WP) and a polarizer P in front of D2 and performed quantum state tomography for four input qubits $|H\rangle, |V\rangle, |D\rangle=\frac{1}{\sqrt{2}}(|H\rangle+|V\rangle)$ and $|R\rangle=\frac{1}{\sqrt{2}}(|H\rangle-i|V\rangle)$. The results are then used to reconstruct the QPT matrix $\chi_{\textrm{exp}}$.

\begin{figure}[t]
   \includegraphics[width=3.5in]{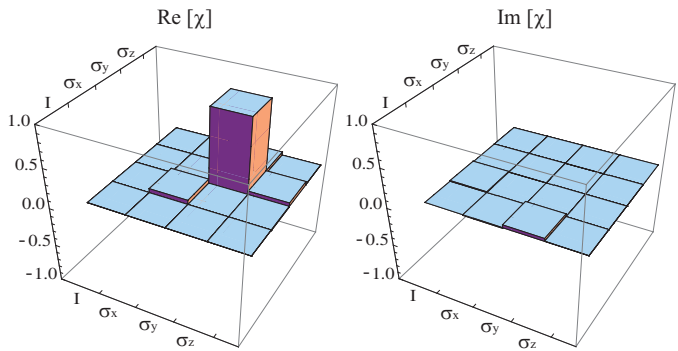}
  \caption{Experimentally reconstructed QPT matrix $\chi_{\textrm{exp}}$ for the $\frac{1}{2}[\sigma_{z},\sigma_{x}]$ operation. It is clear  $[\sigma_{z},\sigma_{x}]$ operation is mainly of the $\sigma_y$ operation. The process fidelity between the experimentally implemented $[\sigma_{z},\sigma_{x}]$ operation and the ideal  $\sigma_{y}$ operation is 0.94.}\label{qpt}
\end{figure}

The reconstructed QPT matrix for the experimentally implemented $\frac{1}{2}[\sigma_{z},\sigma_{x}]$ operation is shown in Fig.~\ref{qpt}. It is evident from the experimental data that the quantum operation we have implemented experimentally is mainly of the $\sigma_y$ operation. The process fidelity, defined as $F=Tr[\chi_{\textrm{exp}}\chi_{\textrm{ideal}}]$, quantifies the quality of the overlap between the experimentally realized operation $\chi_{\textrm{exp}}$ and the desired ideal operation $\chi_{\textrm{ideal}}$. In experiment, the process fidelity of $F=0.94$ has been observed with $\chi_{\textrm{ideal}}$ corresponding to the ideal $\sigma_y$ operation.

So far, we have determined that the commutator $[\sigma_z, \sigma_x]$ is proportional to $\sigma_y$ and the proportionality constant $k$ is yet to be determined. Let us now discuss how to determine $|k|$. For the input qubit $|\psi_0\rangle$, the single-photon transmitted at the first beam splitter (just before the second beam splitter) has the quantum state $\sigma_z \sigma_x |\psi_0\rangle/\sqrt{2} = i \sigma_y |\psi_0\rangle/\sqrt{2}$. Similarly, the single-photon reflected at the first beam splitter (just before the second beam splitter) has the  state $i\sigma_x \sigma_z |\psi_0\rangle/\sqrt{2} = \sigma_y |\psi_0\rangle/\sqrt{2}$. To determine $|k|$, we first record the rate of events at D2, $N$, for the commutator operation to the input state, $\frac{1}{2}[\sigma_{z},\sigma_{x}]|\psi_0\rangle$. We then block the lower path of the interferometer to record the rate of events at D2, $N_u$, corresponding to $\sigma_y|\psi_0\rangle/{2}$, and block the upper path to record the rate of events at D2, $N_l$, corresponding to $i \sigma_y|\psi_0\rangle/{2}$. Since $|k| = |[\sigma_z,\sigma_x]/\sigma_y|$, $|k|$ can be determined from the measured rates using the relation $|k|=N/(N_u+N_l)$.  The experimental obtained value is $|k|=2.12\pm0.18$.


Determining the phase factor of $\pi/2$ in $k$ requires additional phase sensitive measurement and, therefore, is more difficult than measuring $|k|$. For this measurement, it is necessary to first interfere the single-photon state at D2 output port, $[\sigma_z,\sigma_x]|\psi_0\rangle=k\sigma_y|\psi_0\rangle$, with a single-photon field of known polarization, $\sigma_y|\psi_0\rangle$. If we then replace the quantum operation $[\sigma_z,\sigma_x]$  with a single $\sigma_y$ operation, the phase difference between the two operations (coming from $\pi/2$ phase factor in $k$) will cause the interferometric signal to change and we can use this result to determine the phase factor in $k$. In our setup, this scheme could in principle be implemented by introducing an additional beam splitter to split the single-photon beam just before the first beam splitter and interfering it with the output of the commutator output port of the original Mach-Zehnder interferometer. In this experiment, however, because of the difficulties in assuring the interferometric stability for a larger interferometer and experimentally implementing replacement of the quantum operation $[\sigma_z,\sigma_x]$ with a singl $\sigma_y$ operation, we have not been able to directly demonstrate the $\pi/2$ phase factor in $k$.


In summary, we have experimentally verified the commutation relation for Pauli spin operators using quantum interference of the single-photon polarization qubit. We have demonstrated that, via single-photon quantum interference, there exists the relative phase factor of $\pi$ between the quantum operations $\sigma_{z}\sigma_{x}$ and $\sigma_{x}\sigma_{z}$. Also, by superposing these operations, we have implemented the commutator, $[\sigma_{z}, \sigma_{x}]$, and the anticommutator, $\{\sigma_{z}, \sigma_{x}\}$, operations on the single-photon polarization qubit. Furthermore, we have demonstrated experimentally that the quantum operation $[\sigma_{z}, \sigma_{x}]$ is predominantly a $\sigma_y$ operation with the profess fidelity of 0.94. Finally, we interferometrically determined that $|k|=2.12\pm0.18$ for the commutation relation $[\sigma_{z}, \sigma_{x}]=k\sigma_y$. Although we have not been able to determine the phase factor in $k$ in this experiment, we have proposed a scheme for such measurement.


This work was supported, in part, by the National Research Foundation of Korea (2009-0070668 and 2009-0084473) and the Ministry of Knowledge and Economy of Korea through the Ultrafast Quantum Beam Facility Program. YSK acknowledges the support of the Korea Research Foundation (KRF-2007-511-C00004).

\textit{Note added.}$-$After completing this manuscript, we have become aware of a related work \cite{yao10}.


\end{document}